\documentclass[singlecolumn, aps,prl,superscriptaddress, showpacs, preprint]{revtex4-1}
\usepackage{graphicx}

\usepackage{color}
\usepackage{threeparttable}

\begin{document}

\title{Low-Energy Optical Phonon Modes in the Caged Compound LaRu$_2$Zn$_{20}$}

\author{K. Wakiya\email[]{Your e-mail address}}
\affiliation{Department of Quantum Matter, Graduate School of Advanced Sciences of Matter, Hiroshima University, Higashi-Hiroshima 739-8530, Japan} 
\author{T. Onimaru} 
\affiliation{Department of Quantum Matter, Graduate School of Advanced Sciences of Matter, Hiroshima University, Higashi-Hiroshima 739-8530, Japan} 
\author{S. Tsutsui} 
\affiliation{Japan Synchrotron Radiation Research Institute, SPring-8, Sayo, Hyogo, 679-5198, Japan}
\affiliation{Institute for Advanced Materials Research, Hiroshima University, Higashi-Hiroshima 739-8530, Japan}
\author{T. Hasegawa} 
\affiliation{Graduate School of Integrated Arts and Sciences, Hiroshima University, Higashi-hiroshima, 739-8521, Japan}
\author{K. T. Matsumoto} 
\affiliation{Department of Quantum Matter, Graduate School of Advanced Sciences of Matter, Hiroshima University, Higashi-Hiroshima 739-8530, Japan} 
\author{N. Nagasawa} 
\affiliation{Department of Quantum Matter, Graduate School of Advanced Sciences of Matter, Hiroshima University, Higashi-Hiroshima 739-8530, Japan} 
\author{A. Q. R. Baron} 
\affiliation{Materials Dynamics Laboratory, RIKEN SPring-8 Center, Sayo, Hyogo 679-5148, Japan}
\author{N. Ogita} 
\affiliation{Graduate School of Integrated Arts and Sciences, Hiroshima University, Higashi-hiroshima, 739-8521, Japan}
\author{M. Udagawa} 
\affiliation{Institute for Advanced Materials Research, Hiroshima University, Higashi-Hiroshima 739-8530, Japan} 
\affiliation{Graduate School of Integrated Arts and Sciences, Hiroshima University, Higashi-hiroshima, 739-8521, Japan} 
\author{T. Takabatake} 
\affiliation{Department of Quantum Matter, Graduate School of Advanced Sciences of Matter, Hiroshima University, Higashi-Hiroshima 739-8530, Japan} 
\affiliation{Institute for Advanced Materials Research, Hiroshima University, Higashi-Hiroshima 739-8530, Japan} 
\date{\today}

\begin{abstract}
The caged compound LaRu$_2$Zn$_{20}$ exhibits a structural transition at $T_{\rm S}$ =150 K, 
whose driving mechanism remains elusive. 
We have investigated atomic dynamics
by the measurements of specific heat $C$ and inelastic X-ray scattering (IXS). 
The lattice part of the specific heat $C_{\rm lat}$ 
divided by $T^3$, $C_{\rm lat}$/$T^{3}$, shows a broad peak at around 15 K, which is reproduced by two Einstein modes with characteristic temperatures of ${\it \theta}_{\rm E1}$=35 K and ${\it \theta}_{\rm E2}$=82 K, respectively. 
IXS measurements along the [111] and [110] directions reveal 
optical phonon modes at 3 meV (35 K) and 7 meV (80 K), respectively,      
whose values agree with the values of $\theta_{\rm E}$'s.  
The first principles calculation has assigned the phonon modes  
at 3 meV as  
the optical  modes of  Zn atoms located at 
the middle of two La atoms. 
The low-energy vibration of the Zn atom perpendicular to the three-fold axis is thought to lead the structural instability of LaRu$_2$Zn$_{20}$. 
\end{abstract}


\maketitle
\maketitle
\newpage
Caged compounds such as intermetallic clathrates, filled skutterudites, and $\beta$-pyrochlore oxides have attracted much attention in recent years, 
since a variety of interesting physical
phenomena arise from large-amplitude and anharmonic vibration of guest atoms inside the cages \cite{reviewcla, reviewsku, beta1}.  
The low-energy optical phonons in the intermetallic clathrates and filled skutterudites reduce the thermal conductivity and thus enhance the thermoelectric figure of merit \cite{glasstc, glasstc2, clathrate}. 
Furthermore, such guest phonon modes  
in rare-earth filled-skutterudites 
are thought to play an important role in 
strongly correlated electronic phenomena such as 
heavy fermion state, superconductivity, and quadrupolar fluctuations 
\cite{Goto, Sanada, Hattori}. 

A family of $R$$T$$_{2}$$X$$_{20}$ ($R$: Rare earth, $T$: Transition metal, $X$=Al and Zn) is  
another class of caged compounds~\cite{structure}. 
They show various physical properties  
such as structural transition, superconductivity, heavy fermion state, and multipolar ordering \cite{YbCo, Gd, YFe, 1, 4, 2, 3, 14, Matsubayashi, Tsujimoto}. 
For example, 
YbCo$_2$Zn$_{20}$ has the largest electronic specific heat coefficient of 8 J/K$^2$ mol among the Yb-based compounds~\cite{YbCo}. 
YFe$_2$Zn$_{20}$ and LuFe$_2$Zn$_{20}$ exhibit behavior of nearly ferromagnetic Fermi-liquid\cite{Gd, YFe}. 
PrIr$_2$Zn$_{20}$ has a non-Kramers $\Gamma_3$ doublet ground state of 4$f^2$ configuration, 
in which a superconducting transition occurs in the presence of antiferroquadrupole order~\cite{1, 4, 2}. 
The coexistence of the superconductivity and quadrupole order has been observed in the isostructural 
Pr compounds PrRh$_2$Zn$_{20}$ and Pr$T_2$Al$_{20}$ ($T$=Ti and V)~\cite{3, 14, Matsubayashi, Tsujimoto}. 
La$T_2$Zn$_{20}$ ($T$=Ru and Ir) and PrRu$_2$Zn$_{20}$ 
undergo structural phase transitions at $T_{\rm s}$=150, 200 and 138 K, respectively~\cite{2, 4}. 

The compounds $RT_2$Zn$_{20}$ crystalize in the cubic CeCr$_{2}$Al$_{20}$-type structure with the space group of Fd$\overline{3}$m. 
The unit cell consists of 8 formula units~\cite{structure}. 
There are five crystallographically different sites; the $R$ atom at 8$a$, $T$ at 16$d$, Zn at 16$c$, 48$f$, and 96$g$ sites.
The atomic displacement parameter of the Zn atom  
at the 16$c$ site, Zn (16$c$), is approximately twice or three times larger than those of other atoms~\cite{structure}. 
Recently, a first principles calculation of La$T_2$Zn$_{20}$ ($T$=Ru and Ir) has pointed out that the Zn (16$c$) vibrates at low frequencies in a two-dimensional plane, which probably induces the structural transition~\cite{6}. 
The Zn (16$c$) is encapsulated in a cage consisting of two $R$ atoms and twelve Zn atoms at the 96$g$ site 
as is shown in the inset of Fig. 1. 
In analogy of intermetallic clathrates and rare-earth filled skutterudites $RT_4$Sb$_{12}$, 
low-energy vibrations of the Zn atoms may  
play a crucial role not only in the structural property but also in the electronic one. 
Whereas the $R$ atom is rattling in $RT_4$Sb$_{12}$~\cite{reviewsku}, the Zn atom in $RT_2$Zn$_{20}$ is vibrating with a large amplitude.  
Experimentally, ultrasonic measurements revealed that 
elastic moduli of
Pr$T_2$Zn$_{20}$ ($T$=Rh and Ir) depend on the ultrasonic frequency at around 2 K \cite{elastic_Rh, 8}, 
as found in some of filled skutterudites \cite{Goto}.  
Furthermore, $^{139}$La-NMR measurements of La$T_2$Zn$_{20}$ ($T$=Ru and Ir) have indicated that Zn (16$c$) moves from the 16$c$ site to off-center positions on cooling below $T_{\rm s}$.
This fact suggests that the low-energy Zn vibration plays a role in the structural transition~\cite{7}.

Specific heat and inelastic X-ray scattering (IXS) measurements 
are powerful techniques to observe low-energy optical phonon modes in solids.   
The specific heat gives information on the phonon density of states, while the IXS measurements on single-crystalline samples provide us with the phonon dispersion relations. 
Combining them, 
the lattice specific heat can be reproduced \cite{RB6}.
In fact, the combinations of specific heat, IXS, and inelastic neutron scattering measurements were usefull to reveal anharmonic vibration of the guest atoms in intermetallic-clathrates, filled-skutterudites, and the $\beta$-pyrochlore oxides~\cite{beta_n, beta_n2, beta, Lee_s, Tsutsui, takaba, 12, Lee_c, 13}.  
   
In the present work, 
we have measured the specific heat and the IXS spectra for LaRu$_2$Zn$_{20}$ 
to observe the low-energy optical phonon modes.   
The phonon modes have been assigned by the first principles calculation. 
Thereby, the optical phonon mode at 3 meV is identified as the optical mode of Zn (16$c$). 
We have reanalyzed the specific heat data of  LaRu$_2$Zn$_{20}$ reported in ref. 35.

Single-crystalline samples of LaRu$_2$Zn$_{20}$ were prepared by the melt-growth method as was described in the previous papers~\cite{YbCo, Gd, YFe}. 
The specific heat was measured by a thermal-relaxation method from 2 to 300 K using a Quantum Design physical property measurement system. 
The high resolution IXS measurements were carried out at BL35XU of SPring-8~\cite{15}. 
We have chosen the set up of Si (11 11 11) backscattering whose    
energy resolution is 1.5 meV. 
The typical $Q$ resolution in the present condition was $\Delta \textbf{\textit Q}$ = (0.12 0.12 0.04) in the reciprocal
lattice unit for LaRu$_2$Zn$_{20}$ with the lattice parameter of 14.4263(2) \AA~\cite{2}. 
By using the sample of 3 mm$^3$ in volume, 
we have measured transverse modes along the [110] direction and longitudinal modes along the [111] direction. 

\begin{figure}[h]
 \begin{center}
 \includegraphics[width=7.2cm,clip]{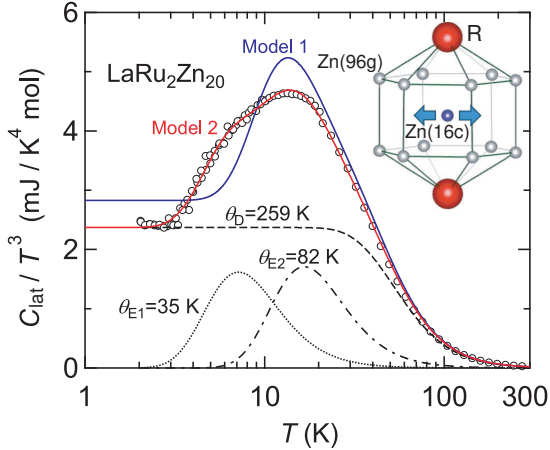}
 \caption{(color online) The specific heat depicted as $C_{\rm lat}$ / $T^3$ vs $T$ for LaRu$_2$Zn$_{20}$. 
The data are taken from ref. 35. 
The solid lines indicate fits using the Debye and Einstein models.  
The dashed, dotted and dashed-dotted curves are individual contributions from the Debye and Einstein oscillators in the model 2. 
The fitting procedure of the Model 2 is described in the text.
}
 \label{f1}
 \end{center}
\end{figure}

\begin{table}[h]
\begin{center}
\caption{Fitting parameters for the lattice specific heat of LaRu$_2$Zn$_{20}$. The labels of D and E denote the Debye and Einstein models, respectively.}
\begin{tabular}{cccc}
\hline \hline 
& Mode & Energy (K) & Number of \\ 
& & & oscillator/f.u.  \\
\hline 
Model 1&D
& 246 & 21+2/3 \\
&E 
& 67 & 4/3 \\  
\hline
Model 2 &D & 259 & 21.1\\
& E1 & 35 & 0.13 \\ 
& E2 & 82 & 1.77 \\
\hline \hline
\end{tabular}
\label{table:LaRu_fitting}
\end{center}
\end{table}

We  show the data of lattice specific heat $C_{\rm lat}$ of LaRu$_2$Zn$_{20}$ in Fig. 1, in which  
$C_{\rm lat}$=$C-\gamma T$ is plotted as $C_{\rm lat}$/$T^{3}$ versus $T$, where $\gamma$ is the Sommerfeld coefficient, 11.9 mJ/K$^2$ mol \cite{JPS}. 
In this plot, the Debye specific heat, which obey  the $T^3$ law, approaches a constant value on cooling below the Debye temperature ${\it \theta}_{\rm D}$. 
The Einstein specific heat, on the other hand, manifests itself 
as a peak with a maximum at $T \approx {\it \theta}_{\rm E}$/5, where ${\it \theta}_{\rm E}$ is the Einstein temperature.  
It is represented as  
\begin{equation}
	C_{\rm E}(T) = 3n_{\rm E}R \left(\frac{{\it \theta}_{\rm E}}{T} \right)^2 \frac{{\rm exp}({\it \theta}_{\rm E}/T)}{[{\rm exp}({\it \theta}_{\rm E}/T)-1]^2},
\end{equation}
where $n_{\rm E}$ is the number of the Einstein oscillators per formula unit. 
As shown in Fig. 1,  $C_{\rm lat}$/$T^{3}$ of LaRu$_2$Zn$_{20}$ exhibits a broad peak at around 15 K, which must result from weak dispersive optical phonon modes. 
The broad peak in $C_{\rm lat}$/$T^3$ is similar to those found in caged compounds such as intermetallic clathrates, filled skutterudites, and $\beta$-pyrochlore oxides~\cite{beta, 12, takaba, 13}. 
The specific heat data of these compounds were 
fit by the sum of 
the Einstein specific heat for the guest atom and the Debye one for the framework atoms. 
To extract the contribution of the low-energy optical modes 
from the specific heat data,  
we used two different procedures for 
least-squares fitting with the Debye and Einstein models, as shown in Table \ref{table:LaRu_fitting}. 

In the Model 1, we assumed that oscillations of Zn (16$c$) are approximated by the Einstein model and the others by the Debye model. 
Because two Zn (16$c$) atoms per formula unit are vibrating with a large amplitude on the plane (see the inset of Fig. 1),   
the number of the Einstein oscillators is set as $n_{\rm E}$=(2/3)$\times$2=4/3 per formula unit. 
Thereby, the total number of the Debye oscillators including the framework oscillations is evaluated as $n_{\rm D}$=23$-$4/3=65/3. 
In this case, fitting parameters are ${\it \theta}_{\rm D}$ and ${\it \theta}_{\rm E}$ solely. 
The fit is shown with the (blue) solid curve in Fig. 1, and 
the parameters are listed in Table \ref{table:LaRu_fitting}.  
Although the peak position is well fit, 
the wide profile below 20 K is poorly reproduced.  

Attempting a better fit to 
the $C_{\rm lat}{/}{T^3}$ data, 
we used the Model 2 with one Debye and two Einstein modes.  
The parameters are the number of oscillators per formula unit ($n_{\rm D}$, $n_{\rm E1}$, and $n_{\rm E2}$) and characteristic temperatures ($\theta_{\rm D}$,  $\theta_{\rm E1}$,  and $\theta_{\rm E2}$) of the Debye and Einstein models, respectively. 
Here, we assume that the total number of atoms per formula unit is 23; $n_{\rm D}$+$n_{\rm E1}$+$n_{\rm E2}$=23.   
The analysis has yielded ${\it \theta}_{\rm D}$=259 K, ${\it \theta}_{\rm E1}$=35 K, and ${\it \theta}_{\rm E2}$=82 K as listed in Table \ref{table:LaRu_fitting}.   
The fit drawn by the (red) solid curve in Fig. \ref{f1} 
well reproduces the broad peak profile. 
The total number of the Einstein oscillators in Model 2, $n_{\rm E1}$+$n_{\rm E2}$=1.9, is larger than $n_{\rm E}$=4/3 in the Model 1 whose value is expected from the two-dimensional vibration of Zn (16$c$) as described above. 
This fact indicates that 
other phonon modes partly contribute to the broad peak. 
On the other hand, $n_{\rm E1}$=0.13 for ${\it \theta}_{\rm E1}$ =35 K in Model 2 is much smaller than $n_{\rm E}$=4/3 for Model 1. 
This discrepancy suggests that  some parts of the low-energy optical phonon modes of Zn (16$c$) are distributed at around 3 meV. 

\begin{figure}[t]
 \begin{center}
 \includegraphics[width=8cm]{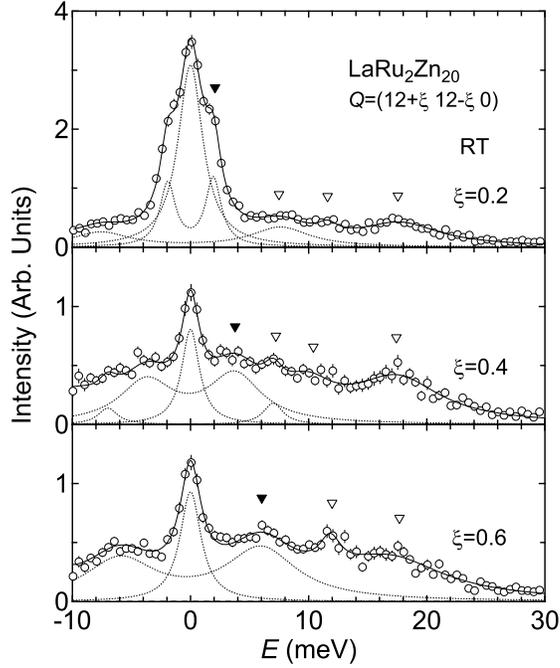}
 \caption{Inelastic X-ray scattering spectra of LaRu$_2$Zn$_{20}$ 
at \textbf{\textit Q}=(12+$\xi$ 12-$\xi$ 0) for $\xi$=0.2, 0.4 and 0.6 at room temperature. 
The solid lines are fits using the Lorentzian function.
Black and white triangles indicate the acoustic and optical phonon excitations, respectively. 
The dashed lines are contributions of the elastic scattering at $E$=0, the acoustic phonon excitation and the optical excitation at 7 meV.}
 \label{f2}
 \end{center}
\end{figure}

\begin{figure}[t]
 \begin{center}
 \includegraphics[width=8cm]{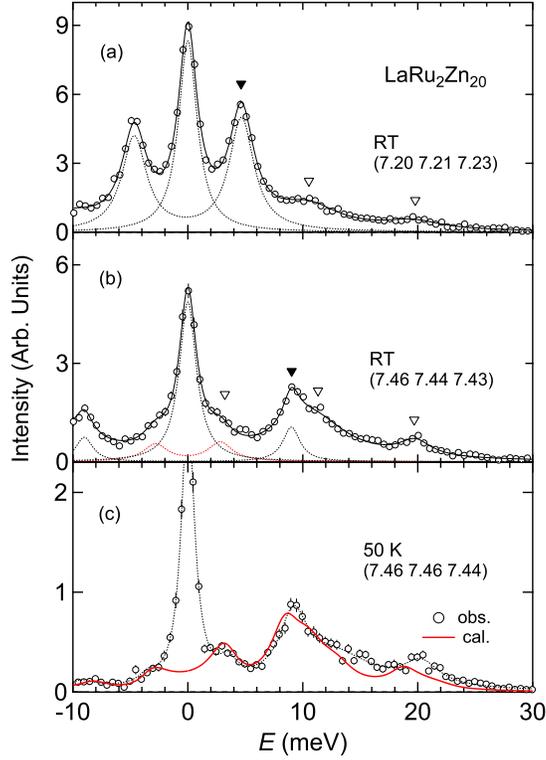}
 \caption{(color online) (a), (b) Inelastic X-ray scattering spectra of LaRu$_2$Zn$_{20}$ in the longitudinal
[111] direction measured at room temperature. The solid lines are fits using the Lorentzian function, while  
black and white triangles indicate the acoustic and optical phonon excitations, respectively. 
The dashed lines are the
individual contributions of elastic scattering at $E$=0, acoustic mode and optical mode at 3 meV. 
(c) IXS spectra measured at 50 K. 
The dashed and solid lines are fit using the Lorentzian function and calculated IXS spectra by the first principles calculations, respectively.}
 \label{f4}
 \end{center}
\end{figure}

\begin{figure}[t]
 \begin{center}
 \includegraphics[width=7.6cm]{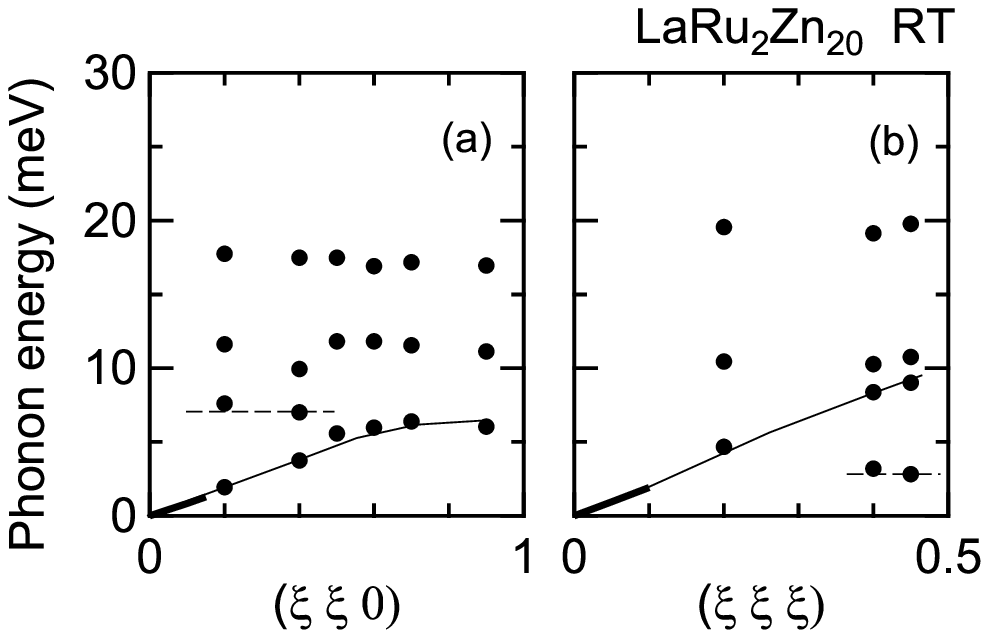}
 \caption{Phonon dispersion curves of LaRu$_2$Zn$_{20}$ measured along (a) \textbf{\textit Q}=(12+$\xi$ 12-$\xi$ 0) and (b) \textbf{\textit Q}=(7+$\xi$ 7+$\xi$ 7+$\xi$) at room temperature. The solid and dashed lines are guides to the eyes. The closed circles represent the phonon peak positions in the inelastic X-ray scattering spectra. 
The error in the energy is comparable to the symbol size. 
The bold lines near the zone center indicate the sound velocity of LaRu$_2$Zn$_{20}$ measured by the ultrasonic measurements
\cite{ishii}. 
}
 \label{f3}
 \end{center}
\end{figure}

Next, we present the results of the IXS measurements of LaRu$_2$Zn$_{20}$.  
Figure 2 shows the room-temperature IXS spectra of LaRu$_2$Zn$_{20}$ at \textbf{\textit Q}=(12+$\xi$ 12-$\xi$ 0) for $\xi$=0.2, 0.4 and 0.6, which correspond to the transverse phonon modes along the [110] direction.  
In addition to the elastic peaks at ${E}{=}$0, there are Stokes and anti-Stokes components of phonon excitations due to phonon creation and annihilation processes, respectively. 
In the stokes part at $\xi$=0.2, a well-defined shoulder exists at around 2 meV which is owing to the acoustic phonon excitation (closed triangle). 
With increasing $\xi$, this peak shifts to higher energy and reaches 7 meV at $\xi$=0.6.  
In addition, a few broad peaks  
exist at 7, 11, and 18 meV (open triangles) which are attributable to weakly dispersive optical phonon modes. 
Because the primitive unit cell of  LaRu$_2$Zn$_{20}$ contains  46 atoms,   
each broad peak would consist of  
several phonon branches. 

Figure 3 (a) and (b) shows the IXS spectra along the \textbf{\textit Q}=(7+$\xi$ 7+$\xi$ 7+$\xi$) for $\xi$=0.2 and 0.45 
which correspond to the longitudinal phonon modes along the [111] direction. 
In the upper panel for $\xi$=0.2, there are an acoustic phonon excitation peak at 5 meV (closed triangle) and two optical phonon peaks at around 10 and 20 meV (open triangles). 
In the middle panel  
for $\xi$=0.45, the acoustic one shifts to 9 meV, and  
a shoulder appears at around 3 meV.  
To confirm whether this shoulder comes from a phonon excitation or not, we have measured the IXS spectrum at lower temperature of 50 K. 
As shown in Fig. \ref{f4} (c), 
a peak is observed at 3 meV in the Stokes part, 
while it is hardly observed in the anti-Stokes part as is expected by the Bose thermal excitation factor. 
Therefore, the peak at 3 meV is identified as the optical phonon mode. 


The above analysis of the IXS spectra of LaRu$_2$Zn$_{20}$ leads to  
the phonon dispersion relations along the [110] and [111] directions 
as shown in Figs. 4 (a) and (b), respectively. 
The closed circles represent the peak positions in the IXS spectra.  
The error in the energy is comparable to the symbol size. 
The solid and dashed lines are guides to the eyes. 
The bold lines in Fig. 4(a) and (b) indicate sound velocities obtained from elastic moduli,  
 $C_{11}$=115.8 GPa, ($C_{11}$-$C_{12}$)/2=31.6 GPa, and $C_{44}$=28.0 GPa, which are measured by ultrasonic technique \cite{ishii}. 
A notable feature in Fig. 4 is the existence of 
low-energy optical phonon branches at 3 and 7 meV. 
The energy values 
agree, respectively, with ${\it \theta}_{\rm E1}$=35 K and ${\it \theta}_{\rm E2}$=82 K estimated from $C_{\rm lat}$. 

To assign the optical phonon modes, we have performed the first principles calculation and compared the results with the observed IXS spectrum. 
The calculated IXS spectrum 
shown with the (red) solid line in Fig. 3 (c)  
agrees with the experimental one. 
According to the calculation, 
the peak at 3 meV is assigned as the Zn (16$c$) vibration, 
whose direction is perpendicular to the three-fold axis connecting the two La atoms as drawn in the inset of Fig. 1. 
We note that the optical phonon mode at 3 meV is also observed in isostructural compound PrRu$_2$Zn$_{20}$ with structural transition at 138 K \cite{16}. 
On the other hand, in PrIr$_2$Zn$_{20}$ with no structural transition, 
the optical phonon mode was not found at 3 meV \cite{16}. 
This fact suggests that the optical mode at 3 meV induces the structural transition in LaRu$_2$Zn$_{20}$. 

In summary, we have observed low-energy optical phonon excitations in LaRu$_2$Zn$_{20}$ by specific heat and  
IXS measurements. 
The temperature dependence of 
$C_{\rm lat}$/$T^{3}$ shows the presence of 
two Einstein modes with characteristic temperatures of 35 and 82 K, respectively. 
The IXS measurements revealed the phonon dispersion relation of the transverse modes along the [110] direction and the longitudinal ones along the [111] direction. 
Along the [111] and [110] directions, optical phonon excitations were found respectively at around 3 and 7 meV, 
whose energy values agree with the Einstein temperatures. 
The first principles calculation allowed us to identify the optical phonon mode at 3 meV as 
the low-energy vibration of Zn (16$c$). 
The absence of 3 meV mode in PrIr$_2$Zn$_{20}$ without showing the structural transition 
supports that the 3 meV mode induces the structural transition at $T_{\rm s}$=150 K in LaRu$_2$Zn$_{20}$. 

\section*{Acknowledgments} 
The IXS experiments were performed under the SPring-8 Budding Researchers Support Program (Proposal No. 2013B1676). 
This work was financially supported by 
MEXT/JSPS KAKENHI Grant Numbers 
 20102004, 20102005, 21102516, 22540345, 23102718, 23740275, and 26707017. 
K. W and K. T. M were supported by JSPS Research Fellowships for Young Scientists.


\begin{thebibliography}{9}
\bibitem{reviewcla} T. Takabatake, K. Suekuni, T. Nakayama, and E. Kaneshita, Rev. Mod. Phys. {\bf86}, 669 (2014). 
\bibitem{reviewsku} H. Sato, H. Sugawara, Y. Aoki, and H. Harima, Handbook of Magnetic Materials (Elsevier, Amsterdam) {\bf18}, 1 (2009). 
\bibitem{beta1} Z. Hiroi, J. Yamaura, and K. Hattori, J. Phys. Soc. Jpn. {\bf81}, 011012 (2012). 
\bibitem{glasstc} G. S. Nolas, J. L. Cohn, G. A. Slack, and S. B. Schujman, Appl. Phys. Lett. {\bf73}, 178 (1998).
\bibitem{glasstc2} B. C. Sales, D. Mandrus, B. C. Chakoumakos, V. Keppens, and J. R. Thompson, Phys. Rev. B. {\bf56}, 15081 (1997).
\bibitem{clathrate} B. C. Sales, D. Mandrus, and R. K. Williams, Science {\bf272}, 1325 (1996).
\bibitem{Sanada} S. Sanada, Y. Aoki, H. Aoki, A. Tsuchiya, D. Kikuchi, H. Sugawara, and H. Sato, J. Phys. Soc. Jpn. {\bf74}, 246 (2005).
\bibitem{Hattori} K. Hattori, Y. Hirayama, and K. Miyake, J. Phys. Soc. Jpn. {\bf74}, 3306 (2005). 
\bibitem{Goto} T. Goto, Y. Nemoto, K. Sakai, T. Yamaguchi, M. Akatsu, T. Yanagisawa, H. Hazama, and K. Onuki, Phys. Rev. B {\bf69}, 180511 (2004).  
\bibitem{structure} T. Nasch, W. Jeitschko, and U. C. Rodewald, Z. Naturforsch. B {\bf52}, 1023 (1997). 
\bibitem{YbCo}M. S. Torikachvili, S. Jia, E. D. Mun, S. T. Hannahs, R. C. Black, W. K. Neils, D. Martien, S. L. Bud${'}$ko, and P. C. Canfield, Proc. Natl. Acad. Sci. U.S.A. {\bf104}, 9960 (2007). 
\bibitem{Gd} S. Jia, N. Ni, G. D. Samolyuk, A. Safa-Sefat, K. Dennis, H. Ko, G. J. Miller, S. L. Bud$'$ko, and P. C. Canfield: Phys. Rev. B. {\bf77}, 104408 (2008). 
\bibitem{YFe} S. Jia, S. L. Bud'ko, G. D. Samolyuk and P. C. Canfield, Nat. Phys. {\bf3}, 334 (2007). 
\bibitem{2} T. Onimaru, K. T. Matsumoto, Y. F. Inoue, K. Umeo, Y. Saiga, Y. Matsushita, R. Tamura, K. Nishimoto, I. Ishii, T. Suzuki, and T. Takabatake, J. Phys. Soc. Jpn. {\bf79}, 033704 (2010).
\bibitem{4} T. Onimaru, K. T. Matsumoto, N. Nagasawa, Y. F. Inoue, K. Umeo, R. Tamura, K. Nishimoto, S. Kittaka, T. Sakakibara, and T. Takabatake, J. Phys.: Condens. Matter {\bf24}, 294207 (2012).   
\bibitem{1} T. Onimaru, K. T. Matsumoto, Y. F. Inoue, K. Umeo, T. Sakakibara, Y. Karaki, M. Kubota, and T. Takabatake, Phys. Rev. Lett. {\bf106}, 177001 (2011). 
\bibitem{14} A. Sakai and S. Nakatsuji, J. Phys. Soc. Jpn. {\bf80}, 063701 (2011). 
\bibitem{Matsubayashi}K. Matsubayashi, T. Tanaka, A. Sakai, S. Nakatsuji, Y. Kubo, and Y. Uwatoko, Phys. Rev. Lett. {\bf109}, 187004 (2012). 
\bibitem{Tsujimoto} M. Tsujimoto, Y. Matsumoto, T. Tomita, A. Sakai, and S. Nakatsuji, Phys. Rev. Lett. {\bf113}, 267001 (2014). 
\bibitem{3} T. Onimaru, N. Nagasawa, K. T. Matsumoto, K. Wakiya, K. Umeo, S. Kittaka, T. Sakakibara, Y. Matsushita, and T. Takabatake, Phys. Rev. B. {\bf86}, 184426 (2012). 
\bibitem{6} T. Hasegawa, N. Ogita, and M. Udagawa, J. Phys.: Conf. Ser. {\bf391}, 012016 (2012). 
\bibitem{elastic_Rh}I. Ishii, H. Muneshige, S. Kamikawa, T. K. Fujita, T. Onimaru, N. Nagasawa,
T. Takabatake, and T. Suzuki, Phys. Rev. B. {\bf87}, 205106 (2013).
\bibitem{8} I. Ishii, H. Muneshige, Y. Suetomi, T. K. Fujita, T. Onimaru, K. T.
Matsumoto, T. Takabatake, K. Araki, M. Akatsu, Y. Nemoto, T. Goto,
and T. Suzuki, J. Phys. Soc. Jpn. {\bf80}, 093601 (2011).  
\bibitem{7} K. Asaki, H. Kotegawa, H. Tou, T. Onimaru, K. T. Matsumoto, Y. F. Inoue, and T. Takabatake, J. Phys. Soc. Jpn. {\bf81}, 023711 (2012).
\bibitem{RB6} S. Kunii, J. M. Effantin, and J. Rossat-Mignod, J. Phys. Soc. Jpn. {\bf66}, 1029 (1997). 
\bibitem{beta_n} H. Mutka, M. M. Koza, and M. R. Johnson, Phys. Rev. B {\bf78}, 104307 (2008).  
\bibitem{beta_n2} K. Sasai, K. Hirota, Y. Nagao, S. Yonezawa, and Z. Hiroi: J. Phys. Soc. Jpn {\bf76}, 104603 (2007).
\bibitem{beta} Y. Nagao, J. Yamaura, H. Ogusu, Y. Okamoto, and Z. Hiroi, J. Phys. Soc. Jpn. {\bf78}, 064702 (2009). 
\bibitem{Lee_s} C. H. Lee, I. Hase, H. Sugawara, H. Yoshizawa, and H. Sato, J. Phys. Soc. Jpn. {\bf75}, 123602 (2012). 
\bibitem{Tsutsui} S. Tsutsui, H. Uchiyama, J. P. Sutter, A. Q. R. Baron, M. Mizumaki, 
N. Kawamura, T. Uruga, H. Sugawara, J. Yamaura, A. Ochiai, 
T. Hasegawa, N. Ogita, M. Udagawa, and H. Sato, Phys. Rev. B. {\bf86}, 195115 (2012). 
\bibitem{takaba} T. Takabatake, E. Matsuoka, S. Narazu, K. Hayashi, S. Morimoto, T. Sasakawa, K. Umeo, and M. Sera, Physica B {\bf383},  93 (2006). 
\bibitem{12} K. Matsuhira, C. Sekine, M. Wakeshima, Y. Hinatsu, T. Namiki, K. Takeda, I. Shirotani, H. Sugawara, D. Kikuchi, and H. Sato, J. Phys. Soc. Jpn. {\bf78}, 124601 (2009). 
\bibitem{Lee_c} C. H. Lee, H. Yoshizawa, M. A. Avila, I. Hase, K. Kihou, and T. Takabatake, J. Phys.: Conf. Ser. {\bf92}, 012169 (2007).
\bibitem{13} M. A. Avila, K. Suekuni, K. Umeo, H. Fukuoka, S. Yamanaka, and T. Takabatake, Phys. Rev. B {\bf74}, 125109 (2006). 
\bibitem{JPS} K. Wakiya, T. Onimaru, S. Tsutsui, K. T. Matsumoto, N. Nagasawa, A. Q. R. Baron, T. Hasegawa, N. Ogita, M. Udagawa, and T. Takabatake, JPS Conf. Proc. {\bf3}, 011068 (2014). 
\bibitem{15} A. Q. R. Baron, Y. Tanaka, S. Goto, K. Takeshita, T. Matsushita, and T. Ishikawa, J. Phys. Chem. Solids {\bf61}, 461 (2000). 
\bibitem{ishii} I. Ishii and T. Suzuki, private communication. 
\bibitem{16} K. Wakiya, T. Onimaru, S. Tsutsui, K. T. Matsumoto,
N. Nagasawa, A. Q. R. Baron, T. Hasegawa, N. Ogita,
M. Udagawa and T. Takabatake, J. Phys.: Conf. Ser. {\bf592}, 012024 (2015).
\end{thebibliography}
\end{document}